\documentclass[journal]{IEEEtran}

\usepackage{graphicx}
\usepackage[caption=false]{subfig}
\usepackage{amsmath}
\usepackage{authblk}
\usepackage[bottom]{footmisc}	
\usepackage{amssymb}
\usepackage{amsthm}
\usepackage[colorlinks, citecolor=black, urlcolor=black]{hyperref}

\usepackage{physics}

\usepackage{braket} 
\usepackage[qm]{qcircuit} 
\usepackage{musicography}  
\usepackage{authblk}
\usepackage{orcidlink}

\usepackage{array}

\begin{document}

\title{\textit{OSC-Qasm}: Interfacing Music Software with Quantum Computing}

\author[1]{Omar Costa Hamido \orcidlink{0000-0001-5077-853X}}
\author[2]{Paulo Vitor Itaboraí \orcidlink{0000-0002-4956-2958}}
\affil[ ]{~\IEEEmembership{ICCMR, University of Plymouth, UK}}
\affil[1]{\textit{omarcostahamido.com/about}}
\affil[2]{\textit{paulo.itaborai@plymouth.ac.uk}}

\markboth{Journal Article, May~2022}%
{Shell \MakeLowercase{\textit{et al.}}: Introducing OSC-Qasm}



\maketitle

\begin{abstract}
\textit{OSC-Qasm} is a cross-platform, Python-based, OSC interface for executing Qasm code. It serves as a simple way to connect creative programming environments like Max (with \textit{The QAC Toolkit}) and Pure Data with real quantum hardware, using the Open Sound Control protocol. In this paper, the authors introduce the context and meaning of developing a tool like this, and what it can offer to creative artists.

\end{abstract}


\IEEEpeerreviewmaketitle

\section{Context}
\label{intro}
\IEEEPARstart{T}{he} use of Quantum Computing (QC) without a Computer Science degree is no longer a far-fetched idea. Recent literature shows that the Arts, Humanities, and Music, have creative researchers exploring both theoretical and practical applications of QC in their respective fields \cite{mirandaQuantumComputingArts2022}.

The development of QC, in itself, is an international and multi-layered effort to articulate both hardware and software in a synergetic dance that allows valid computations to be performed. Very recently, IBM released their expanded roadmap to develop QC. In it, they explain with great detail and enthusiasm the (literal) shape of future quantum processors and when they plan to release them. This description is accompanied with a graph showing the focus areas and layers of this roadmap, including: systems, kernel, algorithms, and models \cite{gambettaExpandingIBMQuantum2022}.

While there seems to be very distinguishable paths into developing quantum hardware (which is enough to tell most QC hardware companies apart), as in the different approaches of a quantum computer (superconducting, photonics, annealing, etc.), the quantum software development is not in the same position yet \cite{gillQuantumComputingTaxonomy2021}. The current main focus on chemistry simulation, machine learning, and noise mitigation algorithms is far from representing the large breadth of QC applications that are currently being explored \cite{montanaroQuantumAlgorithmsOverview2016}\cite{upamaEvolutionQuantumComputing2022}. After all, the purpose of pursuing a universal quantum computer is precisely to be able to compute any type of problem \cite{bernsteinQuantumComplexityTheory1993}.

Without any other incentive to do it otherwise, most quantum software development so far has been pursued independently, and closely connected to different QC programming frameworks. In the majority of cases, it means that researchers are writing Python scripts - even though this may not always be preferable. In the end, they all are dealing with quantum circuits that share a very consensual textbook definition across all layers and fields of development. At this point it becomes relevant to refer to OpenQASM (Open Quantum Assembly Language) as an important effort to unify different QC frameworks in their discrete-variable quantum circuit definition \cite{crossOpenQuantumAssembly2017}.

In the field of Computer Music, in particular, the first author has explored the practical implications of QC for creative practice, focusing on Quantum-computing Aided Composition (QAC) \cite{hamidoAdventuresQuantumland2021}\cite{hamidoQACQuantumcomputingAided2022}. This work has enabled the development of \textit{The QAC Toolkit}, a software toolkit for musicians and artists to build, run, and simulate quantum circuits, using the Max visual programming environment \cite{Max2021}. 

\textit{The QAC Toolkit} represents a shift in the focus of the quantum software development from the QC frameworks and into the Computer Music ecosystem. At the same time, it doesn't diverge entirely - it still allows users to export Qiskit code as well as Qasm \cite{gambettaQiskitQiskitQiskit2021}. One of the main reasons for this, is to still be able to connect with the real quantum hardware that only the original frameworks provide. In \cite{hamidoAdventuresQuantumland2021} this was originally proposed with an object called \texttt{[och.qisjob]}, inspired by the \textit{QisJob} project by Jack Woehr \cite{woehrQisJob2021}. 

In late 2021, with the advent of the QuTune Project and the 1st International Symposium on Quantum Computing and Music Creativity (ISQCMC), we witnessed an increased interest in enabling musical applications to execute quantum circuits \cite{QuTuneProject}\cite{1stInternationalSymposium}. The \textit{OSC-Qasm} project was thus born, as a direct descendant from \texttt{[och.qisjob]}, and an attempt to abstract the required process for executing quantum circuits, in the form of Qasm code, just by simply exchanging OSC messages.\footnote{\textit{OSC-Qasm} software and source code is available for download \cite{HamidoOSCQasmV22022}} 

\subsection{OSC Protocol}
\label{protocol}
The Computer music community has long explored different network communication implementations for exchanging, mapping and controlling musical parameters within and between computers, digital instruments and software. One of the most prominent protocols for this purpose is Open Sound Control (OSC). 

OSC is a protocol based on UDP networks and consists of a strictly formatted binary message. As a consequence, this protocol comprises an inherent Client-Server communication logic. In other words, there is a Server listening to incoming UDP messages (as long as they are formatted according to the OSC convention) and one (or multiple) Clients connecting to it, sending information.

Most importantly, this message is structured in two parts. There is a \textit{path} and a \textit{message} (that can also be a list). The \textit{path} is an arbitrarily chosen word that declares the subject or place to which this message is addressed to - similar to URLs pointing to webpages, or pointing to a file on disk using a directory path. The Server is then capable of parsing and routing the received information according to their respective paths.

\section{OSC-Qasm Server}
\label{server}
The \textit{OSC-Qasm} project includes both a server application and client patches. The \textit{OSC-Qasm Server} is a cross-platform, Python-based, application that runs quantum circuits on quantum backends (both simulators and real hardware) using \textit{Qiskit} \cite{gambettaQiskitQiskitQiskit2021}\cite{QiskitWebpage}. It consists of an OSC server that listens to incoming messages with Qasm scripts from any OSC client. The structure of these messages uses the OSC path \texttt{/QuTune} and a list of 1 to 3 values that include, in order: the Qasm code, the number of shots, and the backend name (see Fig.\ref{fig:qasm}). 

Once \textit{OSC-Qasm Server} receives a valid message, it will automatically transform the Qasm script into a \texttt{qiskit.QuantumCircuit()} object. The circuit is then either simulated in one of the Qiskit simulator backends or executed as an IBMQ job on an available backend \cite{IBMQWebpage}. \footnote{Running jobs on real IBMQ hardware requires user credentials. The number of available backends depends on these credentials. At the time of this writing, several backends are publicly available with a free IBMQ account} Finally, the resulting aggregated counts of the computation are sent back to the client as an OSC message with the path \texttt{/counts}.\footnote{Additional information is also sent back under the \texttt{/info} path, and errors are flagged with the \texttt{/error} path. It is important to note that the receiving client is, in fact, also running an OSC server}

Originally a set of Python scripts, \textit{OSC-Qasm Server} is, as of version 2.0.0, distributed as a standalone application for Linux, MacOS, and Windows. By default, it runs with a graphical user interface (GUI), but it can also be executed as a Command-Line Interface (CLI), provided the necessary arguments are included. Sections \ref{gui} and \ref{cli} will explain in more detail their affordances and differences.

\subsection{GUI}
\label{gui}

\begin{figure}[ht]
\centering
\includegraphics[width=8.5cm]{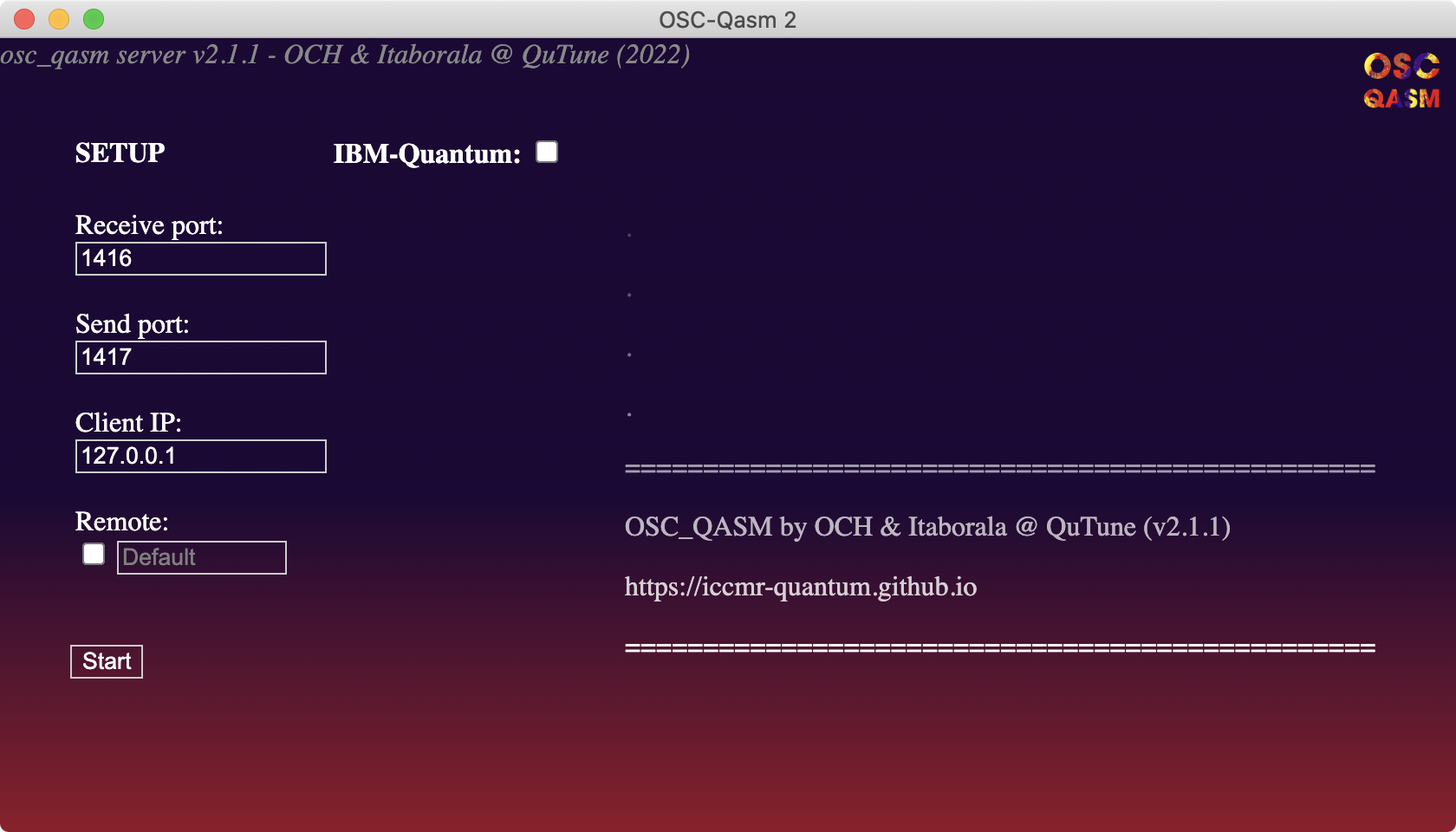}
\caption{OSC-Qasm Graphic User Interface}
\label{fig:gui}
\end{figure}

Intended to be accessible to users who don't code in Python, the Graphic User Interface of \textit{OSC-Qasm server} displays all the customizable parameters on a small window, as depicted in Fig. \ref{fig:gui}. 

On the left side, there is a network setup form with a Start/Stop button at the bottom. These fields are relevant to the configuration of the OSC server. By default, the server is set to run locally (meaning that both server and client will be running on the same device), listening on UDP port 1416 and sending messages back to another application on the same machine (IP 127.0.0.1), listening on port 1417. Optionally, if enabled, the \texttt{Remote} field allows the server to listen to incoming messages from other machines in a network (see section \ref{net}). 

Without further customization, \textit{OSC-Qasm} will use a Qiskit simulator named \textit{qasm\_simulator} to run the quantum circuits. As explained before, a client can request a specific backend by sending its name with the OSC message (see more in section \ref{client}). However, in order for \textit{OSC-Qasm Server} to be able to access real IBM Quantum hardware in the cloud, it is necessary to enable the \texttt{IBM-Quantum} checkbox, that reveals a second column of additional fields. These allow the user to fill the necessary account token, and specify the Hub, Group, and Project details to use.

The right side reveals a console-like monitoring area of up-scrolling text where it displays information about the running processes. 
After changing all the necessary options, the user only needs to click the \texttt{Start} button to boot the OSC server. If the server boots successfully, two lines will appear in the monitoring section, displaying the server arguments and declaring itself ready to exchange OSC messages.

\subsection{CLI}
\label{cli}
In alternative to running the GUI, the \textit{OSC-Qasm Server} can also be executed as a \textit{Command-Line Interface} program from a Terminal or a Command Prompt, using the \texttt{--headless} flag. The default network configurations are identical to the GUI version. The main difference of the CLI mode is that the server boots automatically, as soon as the program is executed, greeting the user with an identical stream of messages as in the GUI monitor section (see Fig. \ref{fig:pop-os}). 

OSC network configurations can be changed with positional arguments on the command-line execution. For instance, the following command launches \textit{OSC-Qasm Server} in headless mode, booting a local server listening to port 3000, and sending the results to a machine with IP 192.168.0.1, on port 3005.\footnote{The user needs to either navigate to the directory where the application is, or drag and drop it onto a terminal window. The Windows version would start with \texttt{OSC\_Qasm\_2.exe}, and the MacOS version with \texttt{OSC\_Qasm\_2.app/Contents/MacOS/OSC\_Qasm\_2}} 

\vspace{5pt}
\footnotesize
\texttt{\$ ./OSC\_Qasm\_2 3000 3005 192.168.0.1 --headless}
\vspace{5pt}

\normalsize
Similarly, additional options such as \texttt{Remote}, and loading IBMQ credentials, are also available using other flag arguments. Additionally, launching \textit{OSC-Qasm} just with the \texttt{--help} flag will list all of the available CLI options.

\subsection{Network Distribution}
\label{net}
A strong advantage of using the OSC protocol is the fact that it is inherently a network protocol. This facilitates more complex mappings and connectivity between media, hence boosting the potential for artistic interactions with quantum computers. The current Client-Server logic allows for a more distributed network setup.

\begin{figure}[ht]
\centering
\includegraphics[width=8.5cm]{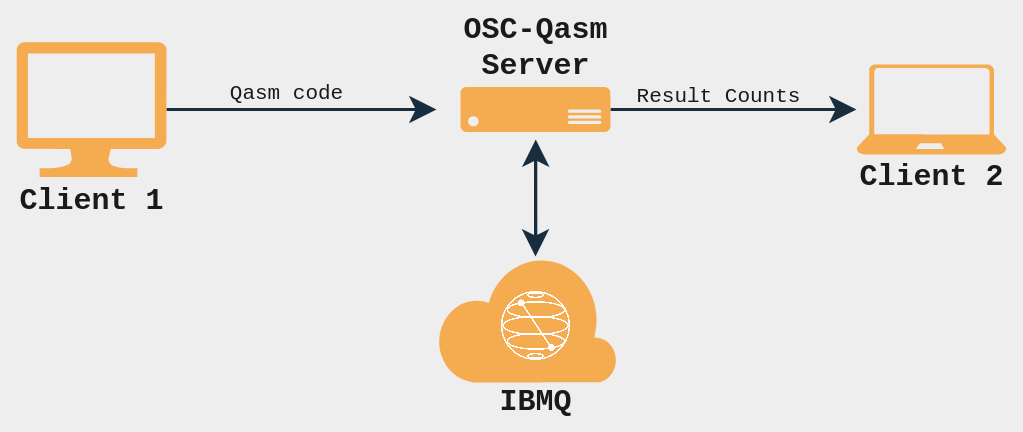}
\caption{A general network distributed \textit{OSC-Qasm}}
\label{fig:network}
\end{figure}

As indicated in Fig. \ref{fig:network}, the \textit{OSC-Qasm Server} can be hosted on a separate machine, as long as it is accessible through a local network or the Web. A dedicated machine for \textit{OSC-Qasm Server} could be useful not just for complex networking but also for offloading the cost of running heavier quantum simulations.

In a possible networked scenario, one client could connect to the server to send Qasm code. \textit{OSC-Qasm} then parses the quantum circuit and runs it on the requested backend. The retrieved computation results could then be sent to a third machine, that will map the results into an ongoing real-time synthesis patch in Max/MSP or SuperCollider (see section \ref{client}).

To allow \textit{OSC-Qasm Server} to be remotely accessible, the \texttt{Remote} option must be enabled. This can be achieved with the checkbox on the GUI, or with the \texttt{--remote} flag in CLI mode. When this option is selected, \textit{OSC-Qasm} will automatically look for the device's main network adapter IP address to host the server. If the server machine is connected to multiple networks (Ethernet, Wifi, VPN, Cellular), the machine's IP address on the desired network can be added, in the field box on the GUI, or as an argument to the flag on the CLI.

\section{OSC-Qasm Clients}
\label{client}
As explained in section \ref{protocol}, the advantage of using an OSC server for receiving Qasm code is that it can come, technically, from \textit{any} OSC-enabled client. More specifically, the authors have explored three Computer Music environments as clients. The only necessary requirement is a correctly formatted OSC message including some valid Qasm code defining a quantum circuit with measurement gates.

The following sections explain each implementation with more detail. The included example patches all share a sample 2-qubit circuit (known as Bell State, see Fig. \ref{fig:qc}). The complete OSC message sent by each client can be seen in Fig.\ref{fig:qasm}.

\begin{figure}[ht]
\texttt{/QuTune "OPENQASM 2.0; \\
include \textbackslash"qelib1.inc\textbackslash"; \\
qreg q[2]; \\
creg c[2]; \\
h q[0]; \\
cx q[0],q[1]; \\
measure q[0] -> c[0]; \\
measure q[1] -> c[1]; \\
" 1024 qasm\_simulator}
    \caption{An example \textit{OSC-Qasm Client} message}
    \label{fig:qasm}
\end{figure}

\begin{figure}[ht]
    \centering
    \scalebox{1.0}{
    \Qcircuit @C=1.0em @R=0.2em @!R { \\
    	 	\nghost{{q}_{0} :  } & \lstick{{q}_{0} :  } & \gate{\mathrm{H}} & \ctrl{1} & \meter & \qw & \qw & \qw\\
    	 	\nghost{{q}_{1} :  } & \lstick{{q}_{1} :  } & \qw & \targ & \qw & \meter & \qw & \qw\\
    	 	\nghost{\mathrm{{c} :  }} & \lstick{\mathrm{{c} :  }} & \lstick{/_{_{2}}} \cw & \cw & \dstick{_{_{\hspace{0.0em}0}}} \cw \ar @{<=} [-2,0] & \dstick{_{_{\hspace{0.0em}1}}} \cw \ar @{<=} [-1,0] & \cw & \cw\\
    \\ }}
    \caption{First Bell state circuit}
    \label{fig:qc}
\end{figure}

\newpage
\subsection{Max}
As explained in Sec. \ref{intro}, \textit{OSC-Qasm} is a direct descendant from the \texttt{[och.qisjob]} Max object. Consequently, Max was the most natural first choice for an \textit{OSC-Qasm Client} \cite{Max2021}. In Max, the user can programmatically create a quantum circuit and retrieve the respective Qasm code definition using \textit{The QAC Toolkit} package \cite{hamidoQACToolkit2021}. Then, a \texttt{[udpsend]} object can be used to send an OSC message to the \textit{OSC-Qasm Server}. Finally, a \texttt{[udpreceive]} object is required in order to receive the computed results back from the server. 

For convenience, and better integration with Max and \textit{The QAC Toolkit} workflow, all the objects related to the message formatting and OSC network exchange were encapsulated in a Max abstraction named \texttt{[osc\_qasm]} (see Fig. \ref{fig:maxoverseas}). Both a help patch and a complete reference page are included with this abstraction.

\subsection{Pure Data}
Another Computer Music environment explored by the authors, with comparable features to Max, is Pure Data (Pd) \cite{puckettePureDataPd}. Unlike Max, however, Pd doesn't include by default objects for formatting OSC messages and a quantum-computing aided composition (QAC) library to programmatically generate quantum circuits. 

\begin{figure}[ht]
\centering
\includegraphics[width=8.5cm]{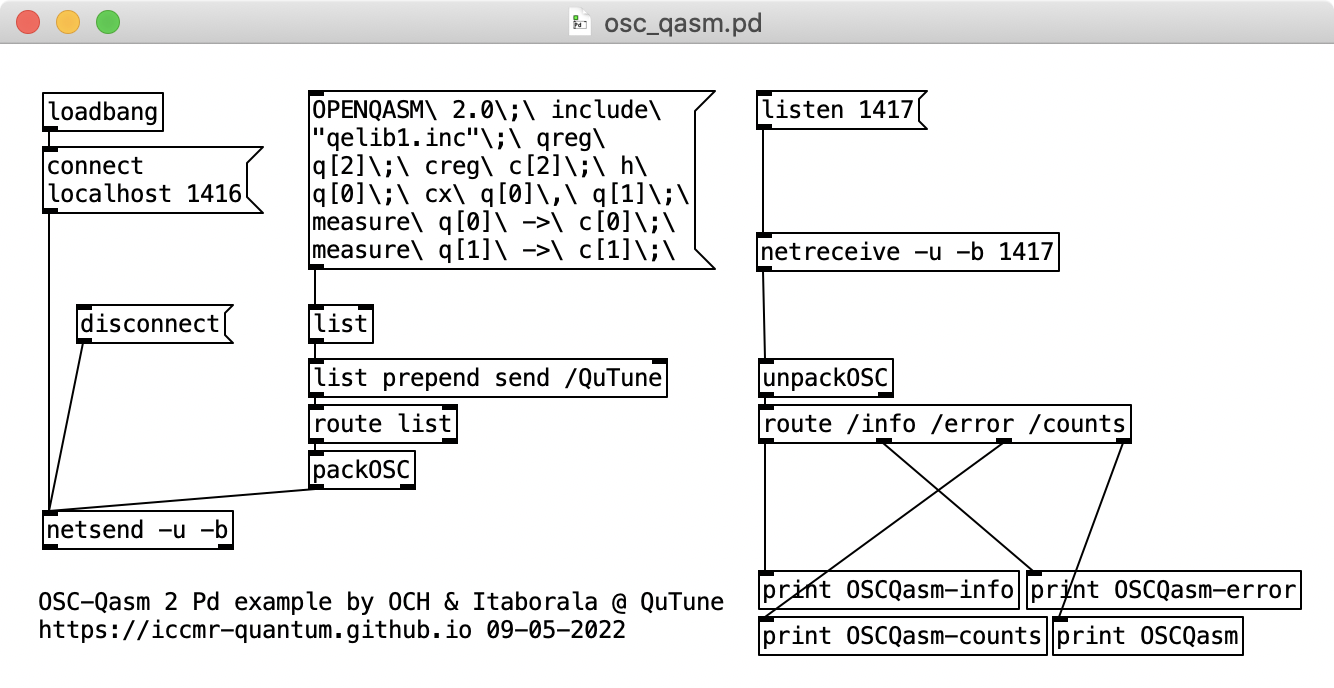}
\caption{\textit{OSC-Qasm Client} in Pd}
\label{fig:pdclient}
\end{figure}

\newpage
To address the OSC message formatting we make use of the \texttt{osc-v0.2} external that can be installed from the included externals library navigator. As for the Qasm code, we decided to include a fixed Qasm string definition in a message object. This Qasm code can be obtained or generated elsewhere. The user must only pay some attention to the Pd message syntax when copying it over. Fig. \ref{fig:pdclient} depicts the Pd example patch that is capable of reproducing the \textit{OSC-Qasm Client} message.

\subsection{SuperCollider}
\begin{figure}[ht]
\centering
\includegraphics[width=8.5cm]{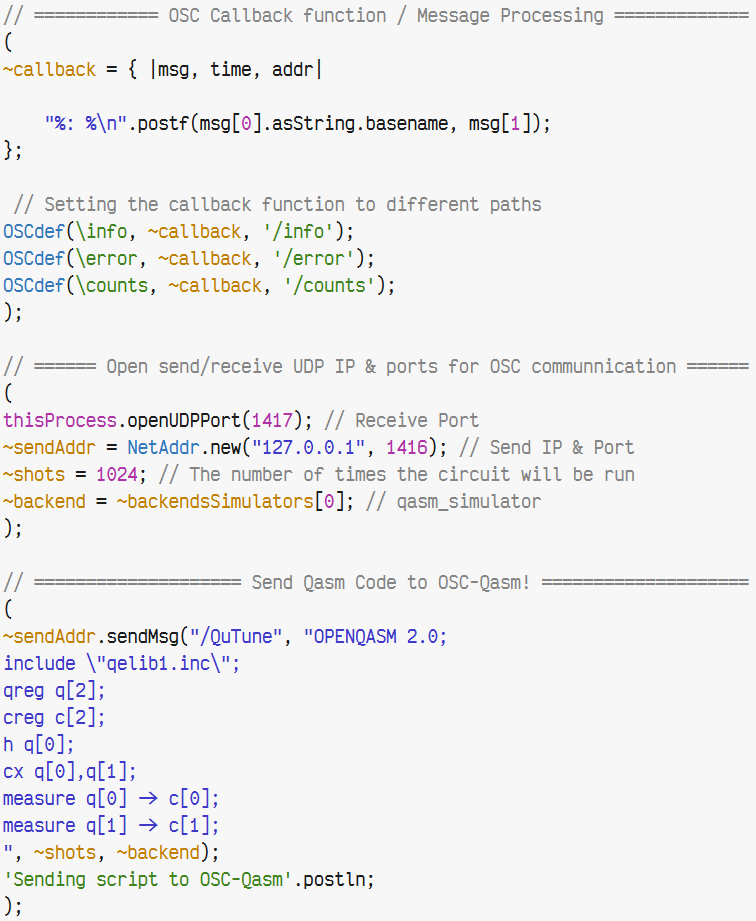}
\caption{OSC-Qasm Client in SuperCollider}
\label{fig:scclient}
\end{figure}

By the same token, the authors have also explored SuperCollider as an \textit{OSC-Qasm Client} \cite{SuperCollider}. Similar to Pd, it does not have a library for \textit{QAC}, however, its text-based paradigm makes it more convenient for writing OpenQASM scripts by hand. In addition, SuperCollider relies strongly on OSC protocol for communications between \texttt{sclang} and \texttt{scsynth}. It has useful functions for defining OSC messaging and parsing structures, for instance, \texttt{NetAddr.sendMsg} and \texttt{OSCdef}. Fig. \ref{fig:scclient} depicts the SuperCollider example client. 

\section{CLOSING REMARKS}
With the release of \textit{OSC-Qasm 2} the authors intend to expand the range of users who are currently working with QC. The \textit{OSC-Qasm} software and source code is publicly available in \cite{HamidoOSCQasmV22022}, with compiled server applications for Linux, MacOS, and Windows, as well as client patches for Max, Pure Data, and SuperCollider. 

The use of OSC protocol, often found in creative programming environments and applications, implies that \textit{OSC-Qasm} can easily be used together with platforms other than those explored by the authors. The attention given to networking capabilities also implies that this software can be used in one or multiple machines, on the LAN or across the internet.

\begin{figure}[ht]
\centering
\includegraphics[width=7.5cm]{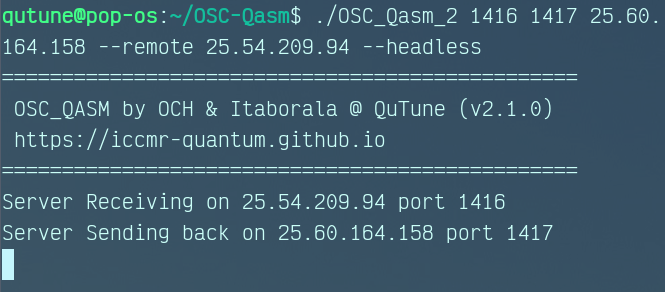}
\caption{\textit{OSC-Qasm Server} used on the distributed network experiment}
\label{fig:pop-os}
\end{figure}

\begin{figure}[ht!]
\centering
\includegraphics[width=8.8cm]{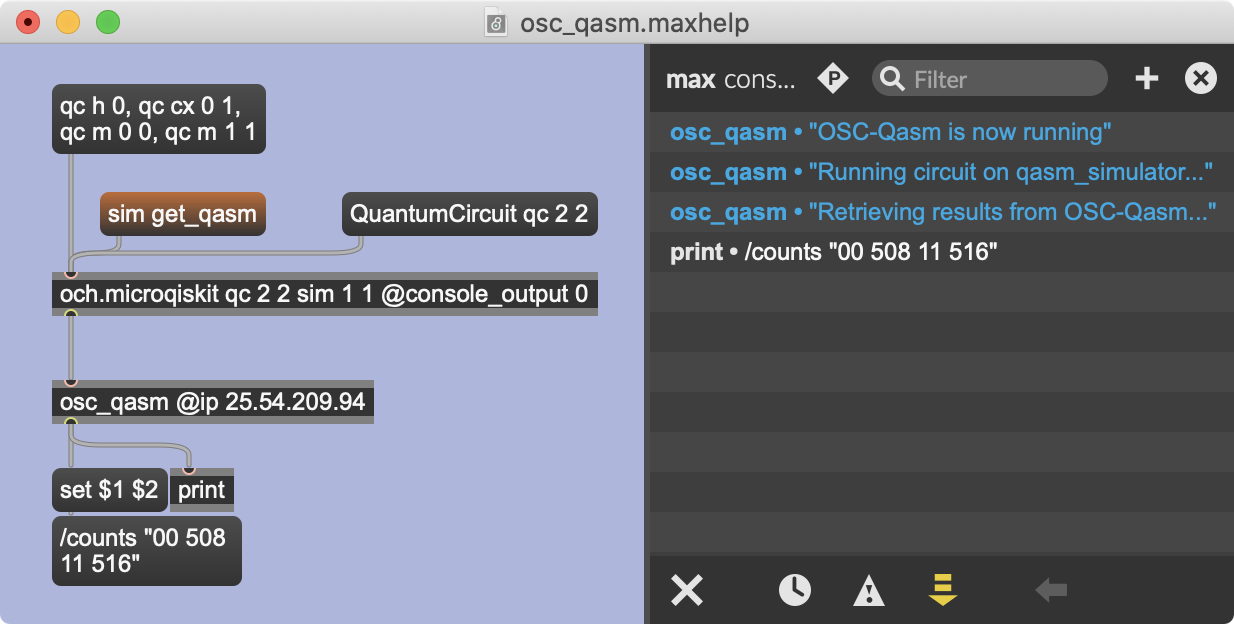}
\caption{Patch used for the distributed network experiment}
\label{fig:maxoverseas}
\end{figure}

In that regard, the authors have already successfully experimented connecting two machines across countries. In this experiment, a Max client was used running on a MacOS laptop, located in Plymouth, UK. The server was running on a Linux (Pop!OS) machine in São Paulo, Brazil (see Figs. \ref{fig:maxoverseas} and \ref{fig:pop-os}). The computers were connected through an Hamachi VPN network \cite{HamachiLogMeIn}.

This tool opens possibilities for multi-user collaborations. The goal is to allow musicians and artists to explore, improve, and make the best out of this tool, by applying QC into their artistic workflows.

\normalsize

\section*{Acknowledgment}
The authors would like to thank other members of the QuTune Project \cite{QuTuneProject}, as well as all the interested participants at the 1st ISQCMC. This work was done thanks to the support from EPSRC QCS Hub.

\bibliographystyle{IEEEtran}
\bibliography{main.bib}



\end{document}